\documentclass[twocolumn,prl,superscriptaddress,showpacs,preprintnumbers,amsmath,amssymb]{revtex4}

\usepackage{graphicx}
\usepackage[usenames]{color}

\newcommand \ket[1]{|#1\rangle}

\begin{document}

\title{Complete control of a matter qubit using a single picosecond laser pulse}
\author{Y.~Kodriano}
\affiliation{Department of physics, The Technion - Israel institute
of technology, Haifa, 32000, Israel}
\author{I.~Schwartz}
\affiliation{Department of physics, The Technion - Israel institute
of technology, Haifa, 32000, Israel}
\author{E.~Poem}
\affiliation{Department of physics, The Technion - Israel institute
of technology, Haifa, 32000, Israel}
\author{Y.~Benny}
\affiliation{Department of physics, The Technion - Israel institute
of technology, Haifa, 32000, Israel}
\author{R.~Presman}
\affiliation{Department of physics, The Technion - Israel institute
of technology, Haifa, 32000, Israel}

\author{T.~A.~Truong}
\affiliation{Materials Department, University of California, Santa
Barbara, California 93106, USA}

\author{P.~M.~Petroff}
\affiliation{Materials Department, University of California, Santa
Barbara, California 93106, USA}

\author{D.~Gershoni}
\affiliation{Department of physics, The Technion - Israel institute
of technology, Haifa, 32000, Israel}
\email{dg@physics.technion.ac.il}

\date{\today}

\begin{abstract}
We demonstrate for the first time that a matter physical two level
system, a qubit, can be fully controlled using one ultrafast
step. We show that the spin state of an optically excited electron,
an exciton, confined in a quantum dot, can be rotated by any
desired angle, about any desired axis, during such a step. For this
we use a single, resonantly tuned, picosecond long, polarized
optical pulse. The polarization of the pulse defines the rotation
axis, while the pulse detuning from a non-degenerate absorption
resonance, defines the magnitude of the rotation angle. We thereby
achieve a high fidelity, universal gate operation, applicable to
other spin systems, using only one short optical pulse. The
operation duration equals the pulse temporal width, orders of
magnitude shorter than the qubit evolution life and coherence times.

\end{abstract}


\maketitle

Matter qubits are essential for any realization of quantum
information processing. Spins of particles are promising
candidates for qubits, since nuclear, atomic or electronic spins are
natural, relatively protected, physical two level systems. Their
spin state can be described as a coherent superposition of the two
levels and thereby geometrically as a vector pointing from the
centre of a unit sphere whose poles are formed by the two levels, to
a point on the sphere surface (Bloch sphere). An important
prerequisite for a qubit is the ability to fully control its state.
A geometrical description of such a universal operation is a
rotation of the qubit's state vector by any desired angle, about any
desired axis~\cite{barenco1995_complete_set, divincenzo2000}.
Naturally, a universal operation should be performed with very high
fidelity and completed within a very short time. The control time should
be orders of magnitude shorter than the qubit's life and decoherence
times~\cite{ladd2010}.

If the two spin eigenstates are non-degenerate (e.g. in a magnetic
field), the spin state evolves in time. This evolution is described
as a precession of the state vector about an axis connecting the
sphere's poles, at a frequency which equals the energy difference
between the two eigenstates divided by the Planck constant.

The control methods demonstrated thus far use a sequence of optical
pulses, which induce fixed rotations of the qubit around axes which
differ from the precession axis (Ramsey interference). A delay
between the pulses allows the qubit to coherently precess between
the pulses and thus a universal operation is achieved. Clearly, such
a sequence of steps increases the time required to perform the
operation, resulting in an operation time comparable to the
precession period. Moreover, the operation fidelity equals the
product of the fidelities of each step. In contrast, we demonstrate
for the first time, that \emph{a single, picosecond}, optical pulse
can be utilized to achieve complete control of a matter qubit,
composed of an optically excited electron (exciton) in \emph{a
single semiconductor quantum dot}~\cite{poem2010, ramsay2008,
delagiroday2010, devasconcellos2010, write_read, poem2011}. Our
demonstration is by no means unique to this technologically
important system and is applicable to other systems as well. We
thereby provide a fast and efficient universal single-qubit gate,
which has not been demonstrated previously in any other platform.

Spins of charge-carriers in semiconductor quantum dots (QDs) are
particularly important candidates for qubits~\cite{loss1998,
imamoglu1999} since they dovetail with contemporary technologies
and since they form an excellent interface between `flying' qubits
(photons) and `anchored' qubits (spins). Spin control in QDs was
demonstrated by radio-frequency pulses~\cite{petta2005, koppens2006}
and by optical means using stimulated Raman
scattering~\cite{berezovsky2008, press2008, degreve2011, greilich2011, godden2012} or by
accumulation of a ``geometrical phase'' through resonant
excitation~\cite{economou2006, economou2007, wu2007, greilich2009,
kim2010, kim2011}.

\begin{figure}[tbp]
  \includegraphics[width=0.48\textwidth]{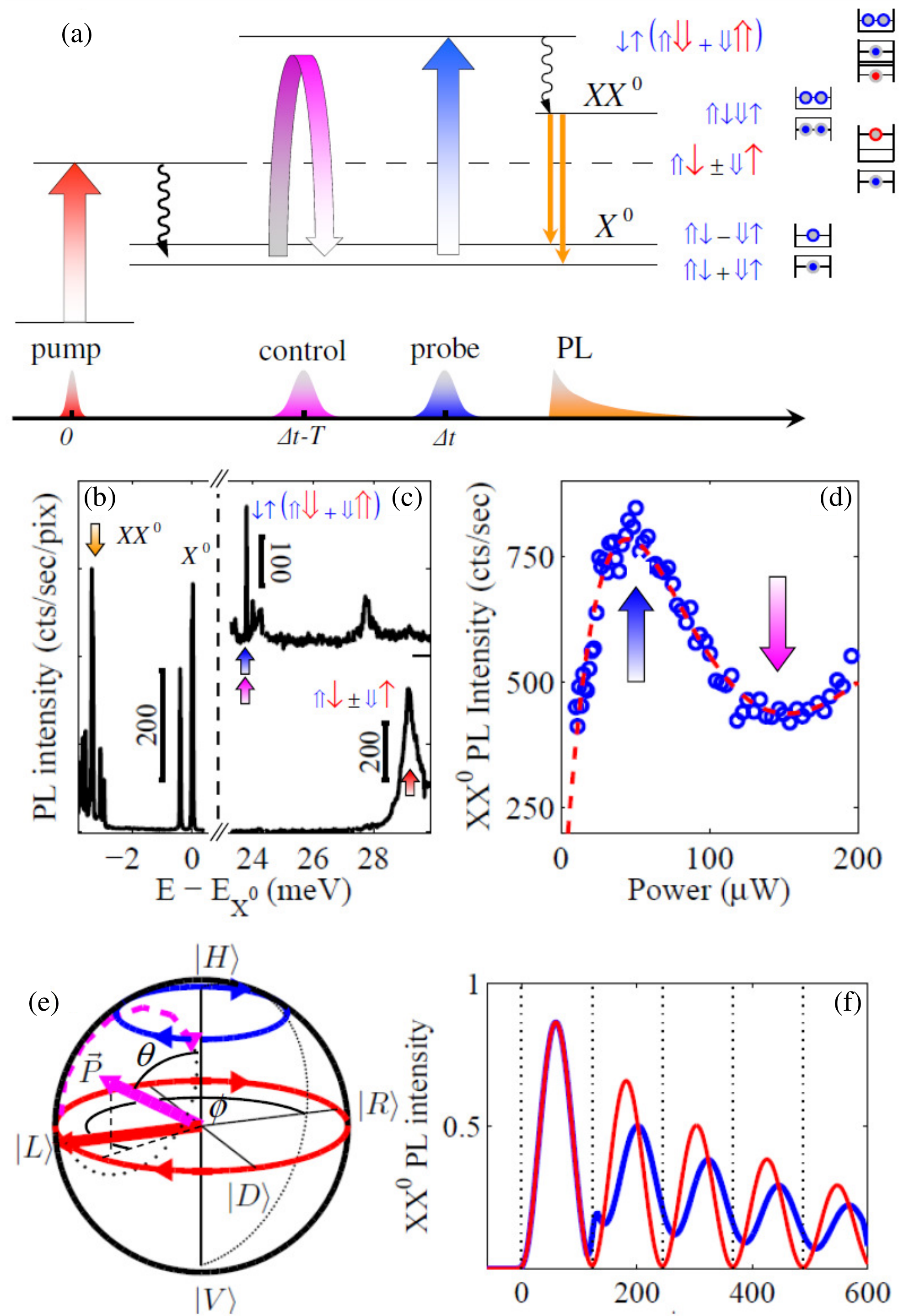}
  \caption{
  Schematic description of the experiment. (a)
The sequence of pulses involved in initialization (pump), control,
and readout (probe) of the exciton's spin state. The relevant
confined exciton and biexciton levels and their spin wavefunctions
are depicted to the right of each level. $\uparrow$ ($\Uparrow$)
denotes electron (hole) spin state and short blue (long red) arrow
denotes ground (excited) single carrier state. (b) PL spectrum of
the QD. (c) PLE spectra of the exciton (bottom) and biexciton (top).
The vertical arrows in (b) and (c) correspond to the optical
transitions denoted in (a) by the same colors. (d). The biexciton PL
intensity vs. the power of the probe laser [blue arrow in (a) and
(c)]. The vertical blue and magenta arrows indicate the laser power
which corresponds to pulse area of $\pi$ (used for the probe) and
2$\pi$ (used for the control) pulses, respectively. The dash line
guides the eye. (e) Bloch sphere representation of the exciton's
spin control. (f) The expected absorption of the probe pulse into
the biexciton resonance as a function of the delay time ($\Delta t$)
between the pump and probe pulses. Blue (red) line describes the
absorption with (without) the control pulse as in (e)} \label{fig:1}
\end{figure}

The qubit discussed here is formed by resonant optical excitation of
an electron from the highest energy full valence band to the lowest
energy empty conduction band. Such excitation can be viewed as
photogeneration of an electron-heavy-hole pair of opposite spins
(bright exciton). Due to the reduced symmetry of the QD, the
exchange interaction within the pair removes the degeneracy between
its two possible spin configurations and forms symmetrical
($\ket{H}$) and anti-symmetrical ($\ket{V}$) eigenstates, which upon
recombination emit light polarized parallel to the major ($H$) and
minor ($V$) axes of the QD, respectively (Fig.~\ref{fig:1}). It
follows that any coherent superposition of these eigenstates,
photogenerated by non-rectilinearly polarized light, precesses in
time with a period $T$, inversely proportional to the energy
difference between the two eigenstates~\cite{write_read}.

The control of the qubit is demonstrated by performing a series of
experiments using sequences of three synchronized optical pulses as
shown in Fig.~\ref{fig:1}(a). The first optical pulse is a polarized
pulse which we tune to an excitonic absorption resonance. It
photogenerates an exciton while translating the light polarization
into exciton spin polarization, with high
fidelity~\cite{write_read}. In order to probe (read) the exciton
spin qubit's state, another picosecond pulse tuned into a
two-exciton (biexciton) resonance~\cite{write_read} is used. This
pulse photogenerates an additional electron-hole pair and transfer
the excitonic population into a biexcitonic
population~\cite{write_read, poem2011, two_photon}. The absorption
of the probe pulse depends on the relative spin orientation of the
two pairs. Since the probe polarization defines the spin of the
second pair, its absorption measures the exciton's spin projection
on its polarization direction~\cite{write_read, poem2011,
two_photon}. The magnitude of the absorption is then directly
deduced from the intensity of the photoluminescence (PL) of the
biexciton emission lines~\cite{two_photon} (Fig.~\ref{fig:1}(b)).

The ability to accurately prepare and probe an excitonic qubit~\cite{write_read} enables us,
 in turn, to demonstrate full control over the qubit state.
For our control operation we use a single 2$\pi$-area optical pulse,
which we tune or slightly detune from a non-degenerate biexcitonic
resonance. This 2$\pi$-pulse transfers the excitonic population into
itself in a process, which involves photon absorption and stimulated
emission. During the process, a relative phase difference is added
between the two eigenstates of the exciton spin, resulting in the
spin vector rotation. We show below that the polarization of the
control pulse determines the spin rotation axis, and the detuning of
the pulse from resonance determines the rotation
angle~\cite{write_read, poem2011}. This is markedly different from
the situation in which a single charge-carrier's spin is controlled
~\cite{petta2005, koppens2006, berezovsky2008, press2008,
degreve2011, economou2006, economou2007, wu2007, greilich2009,
kim2010, kim2011} or a degenerate resonance is used for the rotation
of the exciton-spin~\cite{poem2011}. In these cases, the
polarization is used to distinguish between the degenerate optical
transitions, and the polarization degree of freedom is lost, leaving
only \emph{a fixed, non-tunabale} rotation axis.

The specific biexciton resonance used here for the probe and for the
control contains two excitons with antiparallel spins and different
spatial symmetries (see Fig.~\ref{fig:1}(a) and the supplemental
material). As a result a polarized pulse in such a resonance couples
only to the exciton with the opposite spin state. For example R
polarized pulse couples to $\ket{L}$, $H$ to $\ket{V}$ and $D$ to $\ket{\bar D}$, where $A$
and $\ket{A}$ are the pulse polarization and the corresponding spin state
of the exciton, respectively~\cite{write_read}.

During the pulse, the coupled part of the state acquires a relative
geometric phase shift of $\pi$ radians relative to the uncoupled
state~\cite{economou2006, economou2007}. On the Bloch sphere, this
relative phase acquisition is viewed as a clockwise $\pi$ rotation
of the state's vector about an axis defined by the control pulse
polarization direction~\cite{economou2006, economou2007, poem2011}
(see Supplemental Material).

Angles of rotations which are different than $\pi$ are realized by
detuning from resonance~\cite{economou2006, economou2007, poem2011}.
The induced phase shift (or rotation angle) depends on the detuning
and the pulse shape. For a hyperbolic secant 2$\pi$-pulse of
temporal form $\propto {\rm sech} (\sigma t) e^{i(E_0 -
\Delta)t/\hbar}$, the phase shift ($\delta$), is given by
\begin{equation} \label{eq:rot_2pi}
\delta = \pi - 2 \arctan \left( \frac {\Delta} {\sigma} \right),
\end{equation}
where $E_0$ is the resonance energy, $\sigma$ is the pulse bandwidth
and $\Delta$ is the detuning from resonance~\cite{economou2006,
economou2007, poem2011}. Hence, using both, polarization and
detuning, a universal gate operation is achieved using a
\emph{single} pulse.

The sample and the experimental setup are described in the
Supplemental Material to this work. The experiment is schematically
described in Fig.~\ref{fig:1}(a), where the relevant energy levels,
the resonant optical transitions between them, and the temporal
order of the pulses to these resonances are depicted. The particular
resonances used in the experiment were identified by PL and PL
excitation (PLE) spectroscopy of the QD with one and two resonant
lasers~\cite{two_photon}, as shown in Figs.~\ref{fig:1}(b) and
\ref{fig:1}(c), respectively. The biexciton PL emission intensity
vs. the intensity of the excitation into the biexciton resonance are
depicted in Fig.~\ref{fig:1}(d), where the intensities which
correspond to a $\pi$- and a 2$\pi$-pulse, used for the probe and
for the control, respectively, are marked. We note that under the
2$\pi$-pulse excitation the signal does not go to zero. This results
from the short lifetime of the resonance used for the control pulse.
The resonantly excited hole decays within ~20
psec~\cite{poem2010_prb} to a lower, metastable biexcitonic
state~\cite{kodriano2010}.
As a result,  part of the excitonic population is lost to the
biexciton  during the 9 psec 2$\pi$-pulse. The biexcitonic
population returns incoherently by spontaneous radiative
recombination within about ~600 ps to the exciton state. Since we
probe the exciton only 122 psec after the control pulse the effect
of the incoherent population is a very small decrease in the
visibility.

In Fig.~\ref{fig:1}(e), the exciton's spin Bloch sphere is used to schematically describe the optical control. The red circle on the sphere surface in Fig.~\ref{fig:1}(e) describes the precession of the exciton spin after initialization by an $L$ polarized pulse, represented by a thick red arrow. The thick magenta arrow in Fig.~\ref{fig:1}(e) describes the direction of the polarization $\vec P (\theta, \phi)$ of the control 2$\pi$-pulse, where the polar ($\theta$) and azimuthal ($\phi$) angles are defined in the figure. The dashed magenta line describes the rotation of the exciton state during the control pulse, and the blue line describes the precession of the exciton state \emph{after} the control pulse ends. The curves in Fig.~\ref{fig:1}(f) describe the magnitude of the absorption of the probe pulse vs. the time difference ($\Delta t$) between the generation (pump) and readout (probe) pulses. The red and blue curve correspond to a measurement without  the control pulse and with it, respectively The control pulse is timed exactly one precession period \emph{before} the probe pulse. Thus, the control action is detected a period after it occurs.

\begin{figure}[tbp]
  \centering
  \includegraphics[width=0.48\textwidth, clip=true, trim=0cm 1.2cm 0cm 0cm]{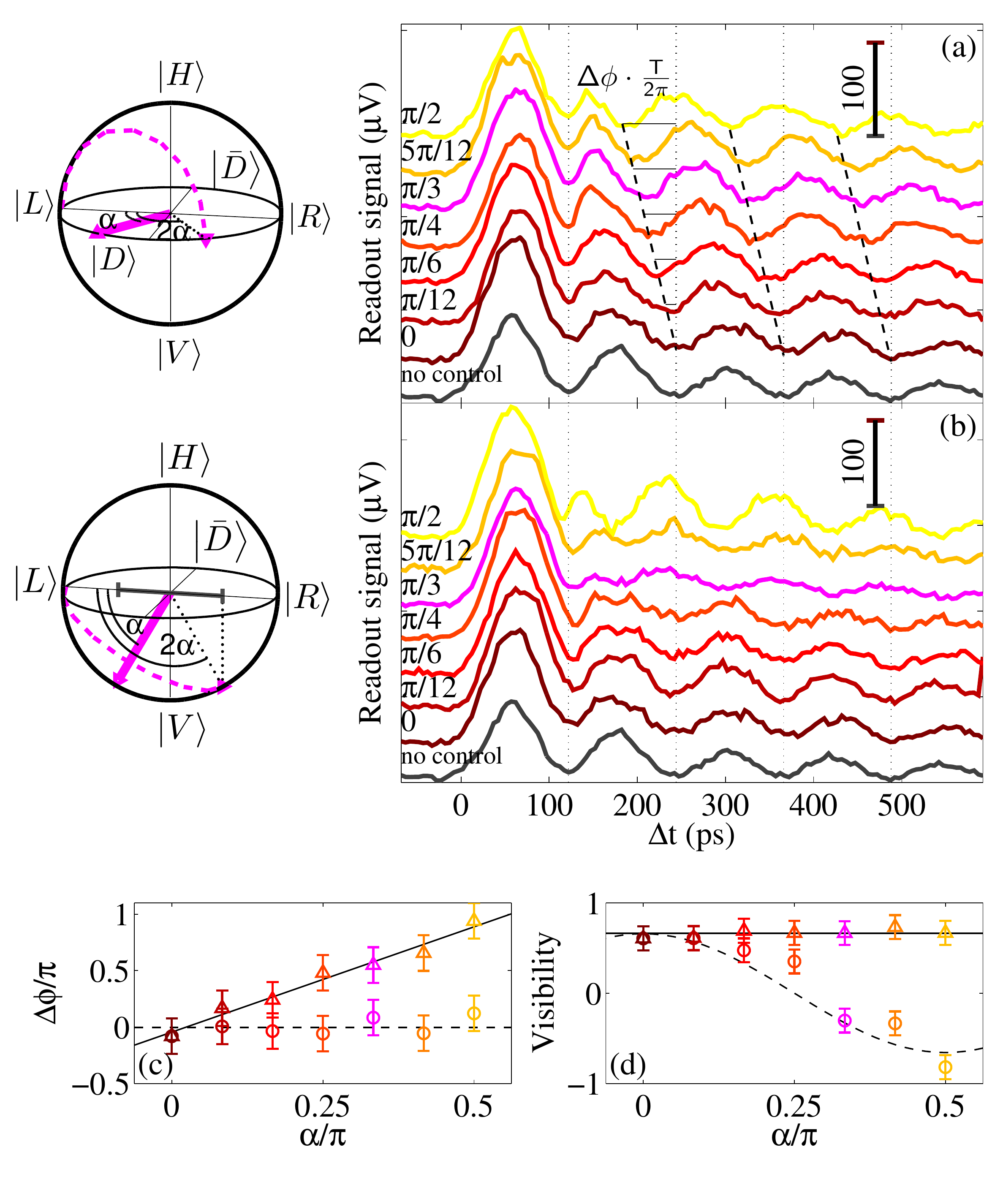}
  \caption{
  Biexciton PL intensity (lock-in detected with the probe pulse) vs. $\Delta$t for various
control pulse polarizations given by $\vec P(\theta, \phi) = \vec P(
\frac{\pi}{2}, \pi + \alpha)$ in (a) and $\vec P(\theta, \phi) =
\vec P( \frac{\pi}{2} + \alpha, \pi)$ in (b) where $\alpha$ varies
from $0$ (L polarization) to  $\pi/2$ [D polarization in (a) and V
polarization in (b)]. The control pulse is one period of oscillation
before the probe. The lowest curve in (a) and in (b) describes the
measurement without the control pulse, and it is used for
normalization. The rotation of the exciton's spin induced by the
polarized control pulse is schematically described as a trajectory
on the Bloch spheres to the left of each panel. (c) [(d)] Symbols:
the phase shifts [normalized visibilities] of the exciton spin
precession induced by the control pulse vs. $\alpha$. The circles'
[triangles'] colors match the colors in (a) [(b)]. The solid
(dashed) lines are best fits to the experimental points in (a)
[(b)].} \label{fig:2}
\end{figure}

\begin{figure}[tbp]
  \centering
  \includegraphics[width=0.48\textwidth, clip=true, trim=0cm 1.2cm 0cm 9cm]{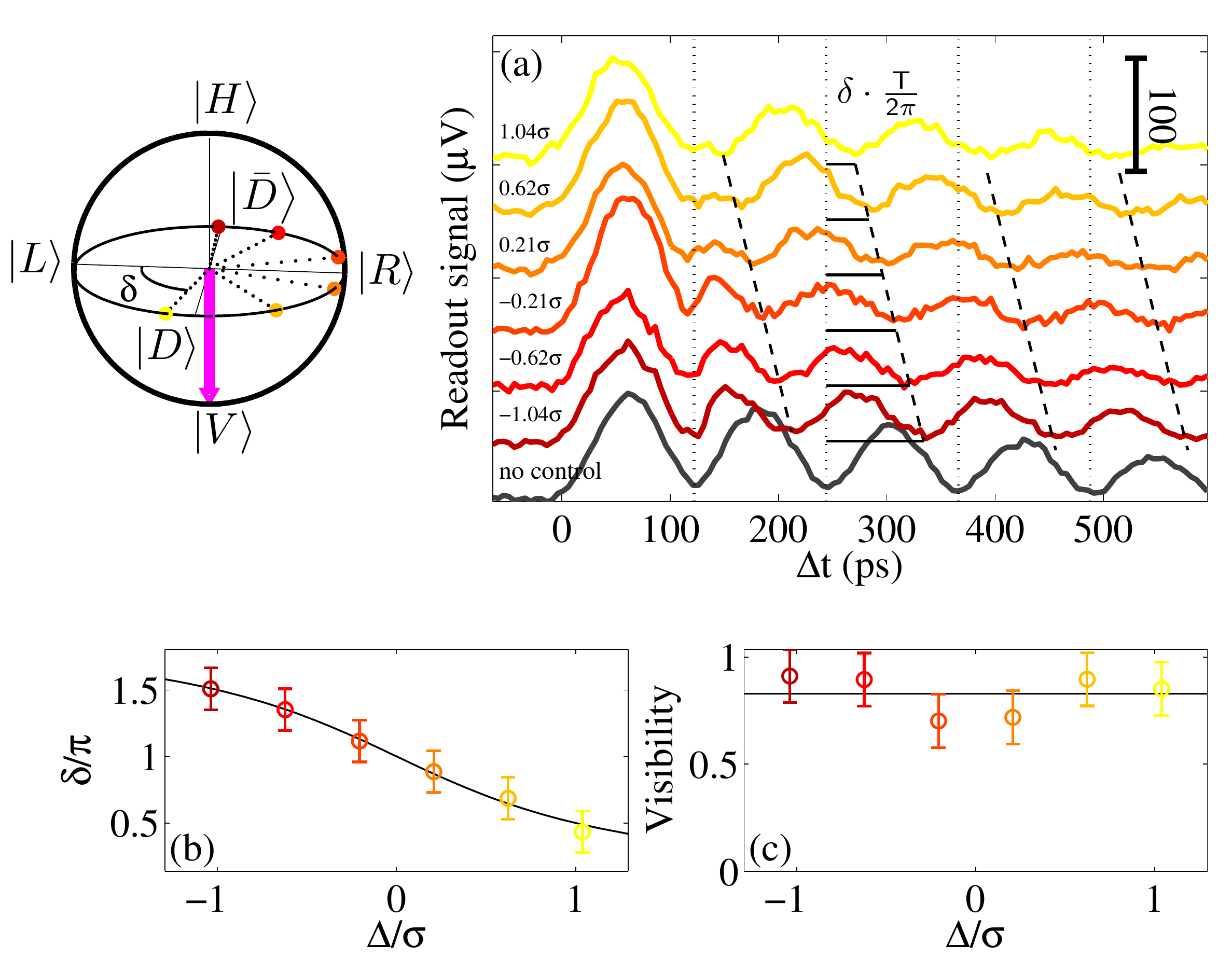}
  \caption{
  (a) Biexciton PL intensity vs. $\Delta$t for a $V$ polarized variably
detuned control pulse. The rotation of the exciton's spin induced by
the control pulse is schematically described as a trajectory on the
Bloch sphere to the left. (b) [(c)] Circles: The phase shifts
(normalized visibilities) of the exciton spin vs. the detuning
$\delta$ in units of the pulse bandwidth $\sigma$. The solid lines
are best fits using Eq.~(\ref{eq:rot_2pi}) in (b) and to a constant
dependence on $\delta/\sigma$ in (c).
  } \label{fig:3}
\end{figure}

In Figs.~\ref{fig:2}-\ref{fig:3} we display three series of experiments which demonstrate our ability to perform universal gate operations on the exciton state using a \emph{single} optical pulse. In these experiments, like in Fig.~\ref{fig:1}, the exciton is always photogenerated in its $\ket{L}$ coherent state by an $L$ polarized pulse. The probe pulse, which in these experiments is delayed continuously relative to the pump, is also $L$ polarized, thus projecting the exciton state onto the $\ket{R}$ state. The lowest, black solid line in each figure describes for comparison the two-pulse experiment in the absence of a control pulse, in which the precession of an $\ket{L}$ photogenerated exciton is probed by the delayed $L$-polarized probe pulse~\cite{write_read}.

The control 2$\pi$-pulse is always launched one precession period
(T=122 psec) \emph{before} the probe pulse. Thus its influence on
the probe signal is noticed only if the control acts after the pump,
i.e at $\Delta t > T$.

In Fig.~\ref{fig:2}, the first two sets of experiments are
presented. Here, the control pulse is tuned to resonance and its
polarization is given by $\vec P (\theta, \phi) = \vec P (\pi/2, \pi
+ \alpha)$ and $\vec P (\theta, \phi) = \vec P (\pi/2 + \alpha,
\pi)$ in (a) and (b), respectively. The angle $\alpha$ spans seven
equally spaced values from 0 ($L$ polarization) to $\pi/2$ ($D$ and
$V$ polarizations in (a) and (b) respectively) as denoted to the
left of each curve. The situation is schematically described on the
Bloch spheres to the left of Figs.~\ref{fig:2}(a) and
\ref{fig:2}(b). The control applies $\pi$ rotation on the exciton
state vector around the polarization direction $\vec P$. The
trajectory on the surface of the Bloch sphere represents this
rotation applied to the initial $\ket{L}$ state. In
Fig.~\ref{fig:2}(a) the rotation always leaves the exciton state on
the equator plane while imparting a phase shift which amounts to
twice the angle $\alpha$ [dashed line in Fig.~\ref{fig:2}(a)]. In
Fig.~\ref{fig:2}(b) the rotations leave the exciton phase fixed
while varying the state projection on the equator (visibility) like
$\cos(2\alpha)$.

In Figs.~\ref{fig:2}(c) and \ref{fig:2}(d) the measured phase shift
of the exciton state and its visibility are displayed, respectively, as a function of $\alpha$.
As expected from the geometrical description, for the rotations in
(a) the visibility does not vary with $\alpha$ while the phase shift
varies linearly from 0 to $\pi$ as $\alpha$ varies from 0 to $\pi/2$
(best fits are described by the solid lines). For the rotations in
(b), the visibility does vary like $\cos(2\alpha)$, while the phase
shift remains unchanged (best fits are described by the dashed
lines).

In order to complete the demonstration of single pulse control, in Fig.~\ref{fig:3}(a) we present a series of measurements in which the control pulse polarization is fixed at $\vec P (\theta, \phi) = \vec P (\pi, \pi)$ ($V$ polarization), while we vary the pulse detuning from resonance. The situation is schematically described on the Bloch sphere to the left (a), which shows the rotations the control pulse imparts to the exciton state. Here, the rotations are achieved through variations in the angle by which the pulse rotates the state around the polarization direction. As can shown in the figure, the exciton state remains on the equator, while it acquires an additional phase of $\delta$ [Eq.~(\ref{eq:rot_2pi})] to its azimuthal angle $\phi$. In Figs.~\ref{fig:3}(c) and \ref{fig:3}(d) the measured phase shift of the exciton state and its visibility, respectively, are displayed as functions of the detuning $\Delta$ measured in units of the the laser bandwidth $\sigma$. While the visibility does not vary with the detuning, $\delta$ varies from approximately $3 \pi/2$ to $\pi/2$ as $\Delta/\sigma$ varies from -1 to 1.

In summary, we demonstrate complete control of the spin state of an exciton in a QD using a single 2$\pi$-area laser pulse, resonant with, or slightly detuned from, a non-degenerate biexciton state. Any desired rotation of the spin state is achieved by controlling three degrees of freedom. The first two are the angles which define the rotation axis around which the exciton spin state is rotated during the short control pulse. These angles are fully determined by the pulse polarization direction. The third is an arbitrary angle of rotation, which is fully determined by the detuning from the biexciton resonance. The method presented here should be applicable to other qubit systems as well. Including, but not limited to, Nitrogen-Vacancy centers in diamonds~\cite{Buckley2010}. The only requirement is an optical transition to a non-degenerate excited resonance.

The support of the US-Israel Binational Science Foundation (BSF), the Israeli Science Foundation (ISF), the Ministry of Science and Technology (MOST), Eranet Nano Science Consortium, and the Technion's RBNI are gratefully acknowledged.


\end{document}


\title{Supplemental material for ``Complete control of a matter qubit using a single picosecond laser pulse''}
\author{Y.~Kodriano}
\affiliation{Department of physics, The Technion - Israel institute
of technology, Haifa, 32000, Israel}
\author{I.~Schwartz}
\affiliation{Department of physics, The Technion - Israel institute
of technology, Haifa, 32000, Israel}
\author{E.~Poem}
\affiliation{Department of physics, The Technion - Israel institute
of technology, Haifa, 32000, Israel}
\author{Y.~Benny}
\affiliation{Department of physics, The Technion - Israel institute
of technology, Haifa, 32000, Israel}
\author{R.~Presman}
\affiliation{Department of physics, The Technion - Israel institute
of technology, Haifa, 32000, Israel}

\author{T.~A.~Truong}
\affiliation{Materials Department, University of California, Santa Barbara, California 93106, USA}

\author{P.~M.~Petroff}
\affiliation{Materials Department, University of California, Santa Barbara, California 93106, USA}

\author{D.~Gershoni}
\affiliation{Department of physics, The Technion - Israel institute
of technology, Haifa, 32000, Israel}
\email{dg@physics.technion.ac.il}





\maketitle

\section{Sample description}
The sample used in this work was grown by molecular-beam epitaxy on a (001) oriented GaAs substrate. One layer of strain-induced $\rm In_xGa_{1-x}As$ quantum dots (QDs) was deposited in the center of a one-wavelength microcavity formed by two unequal stacks of alternating quarter-wavelength layers of AlAs and GaAs, respectively. The height and composition of the QDs were controlled by partially covering the InAs QDs with a 3 nm layer of GaAs and subsequent growth interruption. To improve photon collection efficiency, the microcavity was designed to have a cavity mode which matches the QD emission due to ground-state e-h pair recombinations. During the growth of the QD layer the sample was not rotated, resulting in a gradient in the density of the formed QDs. The estimated QD density in the sample areas that were measured is $\rm 10^8 cm^{-2}$; however, the density of QDs that emit in resonance with the microcavity mode is more than two orders of magnitude lower~\cite{ramon2006}. Thus, single QDs separated by few tens of micrometers were easily located by scanning the sample surface during micro-photoluminescence (PL) measurements. Strong antibunching in intensity autocorrelation measurements was then used to verify that the isolated QDs are single ones and that they form single-photon sources.

\section{Experimental setup}
\begin{figure*}[tbh]
  \includegraphics[width=0.95\textwidth]{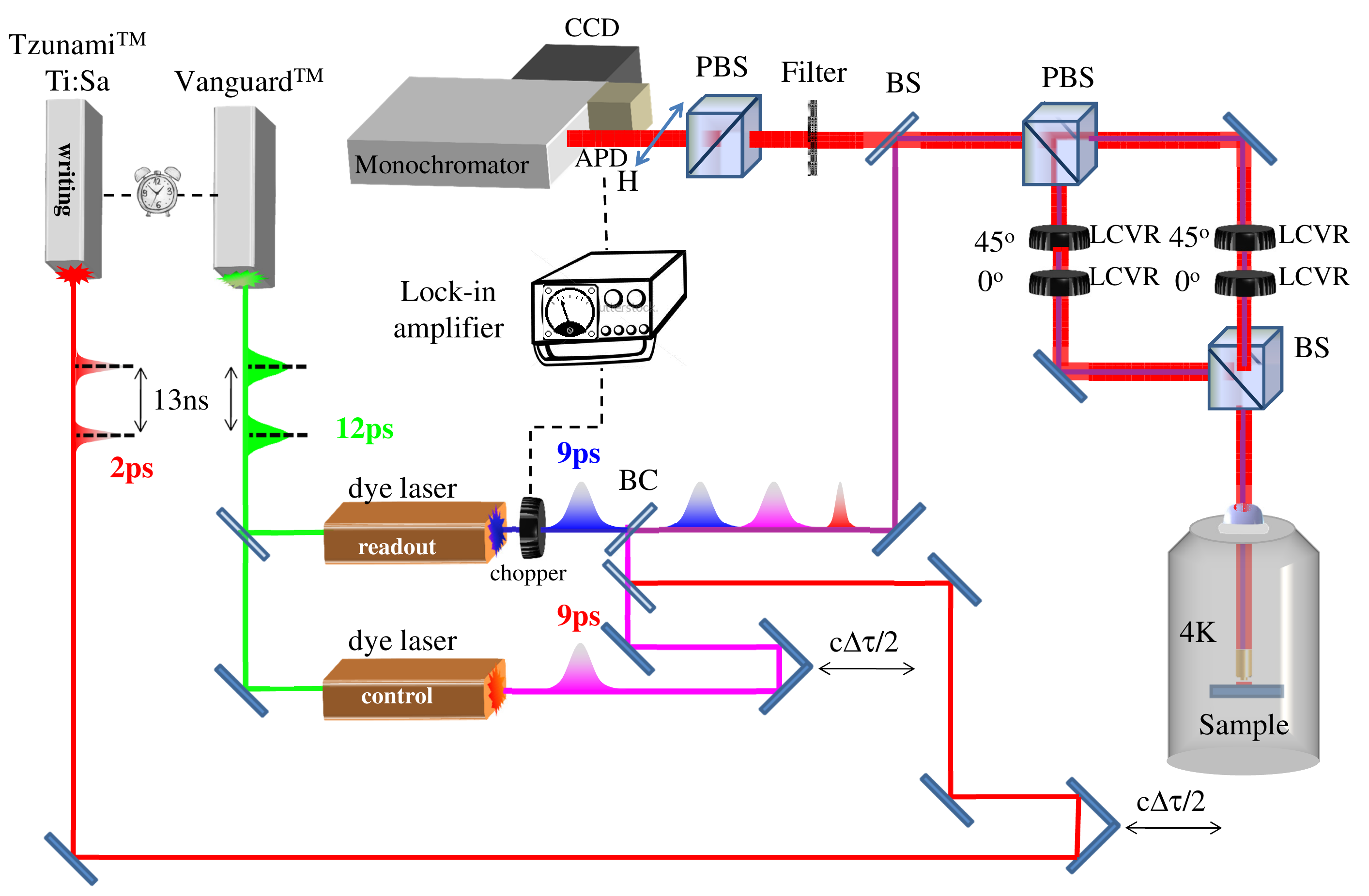}
  \caption{
  Schematic description of the experimental setup. The delay between the three pulses is controlled by 2 computer controlled motorized retroreflectors. (P)BS stands for (polarizing) beam splitter, VBS for a variable beam splitter, BC for beam combiner, and LCVR for a liquid crystal variable retarder.} \label{fig:s1}
\end{figure*}

The experimental setup that we used for the optical measurements is described in Fig.~\ref{fig:s1}. The sample was placed inside a sealed metal tube immersed in liquid helium, maintaining a temperature of 4.2 K. A $\times$60 microscope objective with numerical aperture of 0.85 was placed above the sample and used to focus the light beams on the sample surface and to collect the emitted PL. In the measurements described in Figs.~3-5 of the Letter we used two dye lasers (Styryl 13), synchronously pumped by the same frequency-doubled Nd:YVO$_4$ (Spectra Physics-Vanguard\texttrademark) laser for generating the resonantly tuned optical pulses, as described in the figure. The repetition rate of the setup was 76 MHz, corresponding to a pulse separation of about 13 nsec. The lasers' emission energy could have been continuously tuned using coordinated rotations of two plate birefringent filters and a thin etalon. The temporal width of the dye lasers' pulses was about 9 psec and their spectral width about 100 $\mu$eV. A third, Ti:Sapphire pulsed laser (Spectra Physics, Tsunami\texttrademark) was locked to the clock of the Vanguard\texttrademark laser. Its pulse duration was about 2 psec and spectral width of about 500 $\mu$eV. The delay between the pulses was controlled by 2 retroreflectors on translation stages. The polarizations of the pulses were independently adjusted using a polarized beam splitter (PBS) and two pairs of computer-controlled liquid crystal variable retarders (LCVRs). The polarization of the emitted PL was analyzed by the same LCVRs and PBS. The PL was spectrally analyzed by a 1-meter monochromator and detected by either a silicon avalanche photodetector (using lock-in detection) or by a cooled charged coupled array detector.

\section{Rotation by a detuned pulse to a non- degenerate biexcitonic transition.}
A coherent exciton spin state may be described as a vector on the Bloch sphere:
\begin{multline}
\ket{X(\theta, \phi)} = \cos \left( \frac {\theta} {2} \right) \ket{H} +ie^{i \phi} \sin \left( \frac {\theta} {2} \right) \ket{V} \\
 = \alpha (\theta) \ket{H} + \beta(\theta, \phi) \ket{V}.
\end{multline}
Here $\ket{H}$ and $\ket{V}$ are the two exciton's eigenstates $\rm 1/\sqrt{2} \left[(1e^1)_{-1/2} (1h^1)_{3/2} + (1e^1)_{1/2} (1h^1)_{-3/2} \right]$ and $\rm 1/\sqrt{2} \left[(1e^1)_{-1/2} (1h^1)_{3/2} - (1e^1)_{1/2} (1h^1)_{-3/2}\right]$ respectively. Here the number denotes the energetic order of the confined single carrier level, the letter stands for the carrier type (e-for electron and h for heavy hole), the superscript (either 1 or 2) for the level's occupation and the subscript for the spin projection of same charge carriers~\cite{two_photon}. These non-degenerate eigenstates form the poles of the Bloch sphere (Fig.~2 of the Letter).

Such an exciton is photogenerated by a short optical pulse (whose bandwidth is larger than the energy difference between the two eigen-energies) resonantly tuned to an excitonic transition provided that the pulse polarization is given by:
\begin{equation} \label{P_x}
\vec P_X(\theta, \phi) = \cos \left( \frac{\theta} {2} \right) \hat H + ie^{i\phi} \sin \left( \frac{\theta} {2} \right) \hat V,
\end{equation}
where $\hat H$ and $\hat V$ represent linear polarizations parallel to the major and minor axis of the QD, respectively~\cite{two_photon, write_read}.

This exciton is excited by a second pulse which is resonantly tuned to a {\emph non-degenerate biexciton resonance}. Here the resonance is $\rm (1e^2) (1h^1 4h^1)_{T_0}$, in which the two electrons form a spin singlet in the ground state and the holes, one in the ground  s-like level and one in the  $\rm d_{HH}$- like 4$\rm ^{th}$ level, form a total spin projection zero triplet~\cite{two_photon}. Maximal absorption occurs if the polarization of the second pulse is crossed-polarized with the exciton spin (and therefore with the polarization of the first pulse, if both pulses are simultaneous),
\begin{equation}
\vec P_{XX}(\pi-\theta, \pi+\phi) = \sin \left( \frac{\theta} {2} \right) \hat H - ie^{i\phi} \cos \left( \frac{\theta} {2} \right) \hat V,
\end{equation}
as described in Refs. \onlinecite{two_photon, write_read}. The cross-polarized exciton state:
\begin{multline}
\ket{\bar X (\theta, \phi)} = \ket{X (\pi - \theta, \pi + \phi)} = \\
= \sin \left( \frac {\theta} {2} \right) \ket{H} - ie^{i \phi} \cos \left( \frac {\theta} {2} \right) \ket{V} = \\
= \alpha (\theta) \ket{H} + \beta(\theta, \phi) \ket{V}.
\end{multline}
is completely unaffected by resonant pulses with this polarization, but maximally coupled to "cross polarized" pulses with polarization  described by Eq.~(\ref{P_x}).

When a control, 2$\pi$-pulse with $\vec P_{XX}$ polarization is applied to an arbitrary coherent excitonic state such as $\ket{X (\theta', \phi')}$
\begin{multline}
\ket{X (\theta', \phi')} = \cos \left( \frac {\theta'} {2} \right) \ket{H} +ie^{i \phi'} \sin \left( \frac {\theta'} {2} \right) \ket{V} = \\
= \alpha (\theta') \ket{H} + \beta(\theta', \phi') \ket{V}.
\end{multline}
which can be conveniently expressed also in terms of the coherent states $\ket{X(\theta, \phi)}$ and $\ket{\bar X(\theta, \phi)}$ as:
\begin{equation}\label{x_p_angles}
\ket{X (\theta', \phi')} = \alpha(\theta^p) \ket{X(\theta, \phi)} + \beta(\theta^p, \phi^p) \ket{\bar X(\theta, \phi)}
\end{equation}
In Eq.~(\ref{x_p_angles}) spherical symmetry considerations are used and the angles $\theta^p$ and $\phi^p$ are measured relative to the pulse's polarization direction.

Since the $\vec P_{XX}(\theta, \phi)$ polarized 2$\pi$ pulse couples only to the $\ket{\bar X(\theta, \phi)}$ part of the exciton wavefunction, this part acquires a geometrical phase of $\delta$ (defined by Eq.~(1) in the Letter) relative to the $\ket{X(\theta, \phi)}$ part of the exciton. Therefore after the control 2$\pi$-pulse ends, the new exciton state is given by:
\begin{multline}\label{x_after_control}
\ket{X (\theta'', \phi'')} = \\
= \alpha(\theta^p) \ket{X(\theta, \phi)} + e^{-i\delta} \beta(\theta^p, \phi^p) \ket{\bar X(\theta, \phi)} = \\
\alpha(\theta^p) \ket{X(\theta, \phi)} + \beta(\theta^p, \phi^p - \delta) \ket{\bar X(\theta, \phi)}.
\end{multline}
Thus, the geometrical description of the control pulse action is a clockwise rotation by an angle $\delta$ about an axis connecting the states $\ket{X(\theta, \phi)}$ and $\ket{\bar X(\theta, \phi)}$, parallel to the polarization direction of the control pulse.

\section{Description of the action of the polarized 2$\pi$ control pulse as a universal rotation}
A unit vector in the polarization direction of an exciton $\ket{X(\theta, \phi)}$, which is coupled to a polarized 2$\pi$ control pulse is given by:
\begin{equation}
\hat n = \left( n_x, n_y, n_z \right) = \left( \cos \phi \sin \theta, \sin \phi \sin \theta, \cos \theta \right),
\end{equation}
where the Cartesian axes are chosen such that: $\hat x \equiv \ket{R} = 1/\sqrt{2} \left( \ket{H} + i \ket{V} \right)$, $\hat y \equiv \ket{\bar D} = 1/\sqrt{2} \left( \ket{H} - i \ket{V} \right)$, $\hat z \equiv \ket{H} $. As discussed above, the effect of the control pulse on the exciton wavefunction can be simply viewed as a rotation about $\hat n$ by the angle -$\delta$. Such a rotation in the eigenstates base is described by the operator:
\begin{equation}
R_{\hat n}\left( \delta \right) = \exp \left( i \vec \sigma \cdot \hat n \frac {\delta} {2} \right),
\end{equation}
where $\vec \sigma \equiv \left( \sigma_x, \sigma_y, \sigma_z \right)$ is the Pauli matrix vector. When this operator is applied to an exciton state such as $\ket{X (\theta', \phi')}$ (Eq.~(\ref{x_after_control})), the freedom in choosing$\theta, \phi$ and $\delta$ constructs a universal rotation of the exciton spin polarization.

\section{Estimation of the phase shift and visibility of the control pulse action.}
Each one of the measurements displayed in Figs.~2-3 of the Letter, was fitted to a functional of the form:
\begin{equation}\label{fit}
C \cdot e^{-\frac {t} {\tau}} \left[ 1 - V \cdot \cos \left( \frac{2 \pi} {T} \cdot \Delta t + \Delta \phi \right) \right],
\end{equation}
where $\tau$, T and C are the exciton lifetime, its precession period and an overall normalization factor. The parameters $\tau$ and T are accurately evaluated independently. The parameters V and $\Delta \phi$ are the visibility and the phase shift of the signal, respectively. These are extracted from the fits of Eq.~(\ref{fit}) to each one of the traces in the three pulse experiments presented in Figs.~3-5, for $\Delta t > T$. Here $\Delta \phi$ is measured relative to $\phi = 0$ in the two-pulse experiments (lowest trace in Figs. 3-5). The visibility of the control pulse is then normalized by the visibility of the two-pulse experiments (without the control pulse), which is slightly less than one~\cite{write_read}.

